\documentclass{nature1}
\usepackage{psfig}

\newcommand\degree{\mbox{$^\circ$}}%
\newcommand\arcmin{\mbox{$^\prime$}}%
\newcommand\arcsec{\mbox{$^{\prime\prime}$}}%

\def\simlt{\mathrel{\hbox{\rlap{\hbox{\lower4pt\hbox{$\sim$}}}\hbox{$<$}}}}
\def\simgt{\mathrel{\hbox{\rlap{\hbox{\lower4pt\hbox{$\sim$}}}\hbox{$>$}}}}

\def\ale{\mathrel{\hbox{\rlap{\hbox{\lower4pt\hbox{$\sim$}}}\hbox{$<$}}}}
\def\age{\mathrel{\hbox{\rlap{\hbox{\lower4pt\hbox{$\sim$}}}\hbox{$>$}}}}

\def\spose#1{\hbox to 0pt{#1\hss}}
\newcommand\lsim{\mathrel{\spose{\lower 3pt\hbox{$\mathchar"218$}}
     \raise 2.0pt\hbox{$\mathchar"13C$}}}
\newcommand\gsim{\mathrel{\spose{\lower 3pt\hbox{$\mathchar"218$}}
     \raise 2.0pt\hbox{$\mathchar"13E$}}}

\title{\LARGE \bf A contemporaneous infrared flash\\from a long $\gamma$-ray burst:\\an echo from the central engine}

\author{C. H. Blake$^{1}$,
  J. S. Bloom$^{1,2}$,
  D. L. Starr$^{13}$,
  E. E. Falco$^{3}$,
  M. Skrutskie$^{9}$,
  E. E. Fenimore$^{7}$,
  G. Duch\^ene$^{12}$,
  A. Szentgyorgyi$^{3}$,
  S. Hornstein$^{10}$,
  J. X. Prochaska$^{4}$,
  C. McCabe$^{11}$,
  A. Ghez$^{10}$,
  Q. Konopacky$^{10}$,
  K. Stapelfeldt$^{11}$,
  K. Hurley$^{5}$,
  R. Campbell$^{6}$,
  M. Kassis$^{6}$,
  F. Chaffee$^{6}$,
 N. Gehrels$^{8}$,
 S. Barthelmy$^{8}$,
 J. R. Cummings$^{8}$,
 D. Hullinger$^{8,14}$,
 H. A. Krimm$^{8,15}$,
 C. B. Markwardt$^{8,14}$,
 D. Palmer$^{7}$,
 A. Parsons$^{8}$, 
 K.~McLean$^{7}$, and
 J. Tueller$^{8}$
}

\begin{document}
\maketitle

\begin{affiliations}
\item Harvard College Observatory,
	Cambridge, MA 02138 USA

\item Astronomy Department, 
	University of California at Berkeley, 
        Berkeley, CA 94720 USA

\item Smithsonian Astrophysical Observatory, 
        Cambridge, MA 02138 USA

\item UCO/Lick Observatory,
      Santa Cruz, CA 95064 USA

\item Space Sciences Laboratory, University of California, Berkeley, CA 94720 USA

\item W.M. Keck Observatories, Kamuela, HI 96743 USA

\item Los Alamos National Laboratory, Los Alamos, NM 87545 USA

\item NASA Goddard Space Flight Center, Greenbelt, MD 20771 USA

\item University of Virginia, Department of Astronomy,
         Charlottesville, VA 22904 USA

\item University of California, Los Angeles, Los Angeles, CA 90095 USA

\item Jet Propulsion Laboratory, California Institute of Technology, Pasadena, CA 91109 USA

\item  Laboratoire d'Astrophysique, Observatoire de Grenoble, 38041 Grenoble Cedex 9, France

\item Gemini Observatory, Hilo, HI 96720 USA

\item University of Maryland,  College Park, MD 20742 USA

\item Universities Space Research Association, Columbia, MD 21044 USA

\end{affiliations}

\begin{abstract}
\phantom{}\vskip 0.05cm 
The explosion that results in a cosmic $\gamma$-ray burst (GRB) is
thought to produce emission from two physical processes --- the
activity of the central engine gives rise to the high-energy emission
of the burst through internal shocking\cite{fmn96} and the subsequent
interaction of the flow with the external environment produces
long-wavelength afterglow\cite{mr97a,sp99a,bel02}. While afterglow
observations\cite{vkw00} continue to refine our understanding of GRB
progenitors and relativistic shocks, $\gamma$-ray observations alone
have not yielded a clear picture of the origin of the prompt
emission\cite{bb04a} nor details of the central engine. Only one
concurrent visible-light transient has been found\cite{abb+99} and was
associated with emission from an external shock. Here we report the
discovery of infrared (IR) emission contemporaneous with a GRB,
beginning 7.2 minutes after the onset of GRB\,041219a\cite{gms+04}.
Our robotic telescope acquired 21 images during the active phase of
the burst, yielding the earliest multi-colour observations of any
long-wavelength emission associated with a GRB. Analysis of an initial IR
pulse suggests an origin consistent with internal shocks. This opens a
new possibility to study the central engine of GRBs with ground-based
observations at long wavelengths.
\end{abstract}


\phantom{}\vskip 0.05cm \noindent 

\begin{figure*}[tbh]
\centerline{\psfig{file=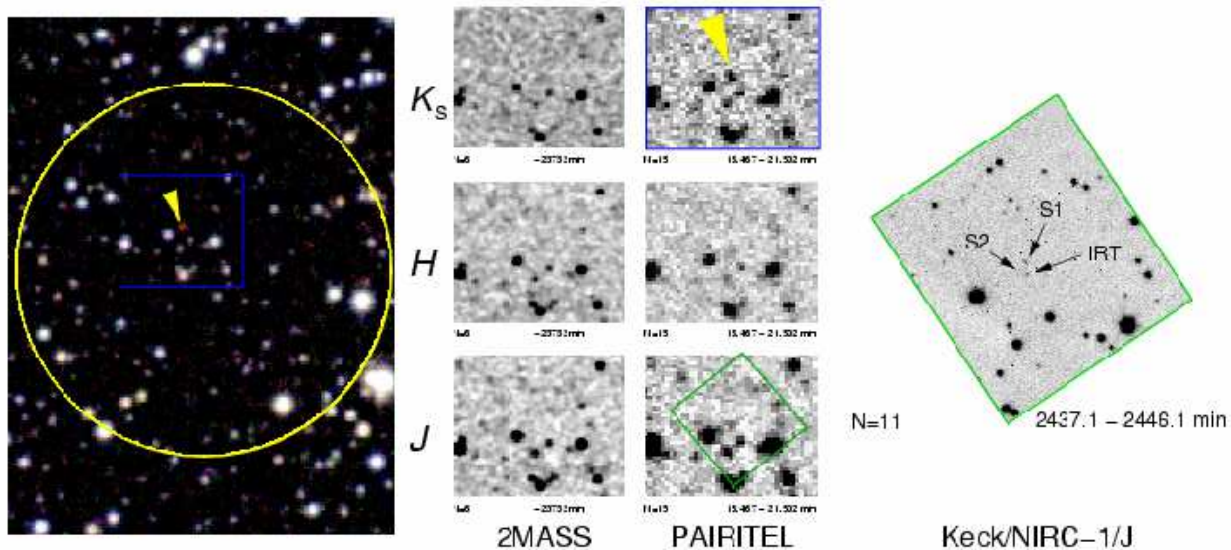,width=7.0in,angle=270}}
\vskip -1in 
\caption[] {\small Images of the IR flash associated with 
GRB\,041219a. The discovery was made with PAIRITEL, located atop
Mt.~Hopkins, Arizona. Normally, the telescope's observing plan is
autonomously scheduled before nightfall by a routine that optimises
the preset priorities and scheduling constraints of all objects in the
PAIRITEL database.  When the new GRB alert was received, a series of
observations were automatically inserted into the observing queue and the
telescope began to slew to the target field. {\bf Left:} False-colour
image of the IR Flash (yellow arrow) of GRB\,041219a inside the
2\arcmin\ (90\% confidence) {\it INTEGRAL} error region\cite{gms+04}
(yellow circle). At early times, the source is the reddest object in
the field, indicative of very high extinction due to dust in the disk
of our Galaxy. {\bf Middle six panels:} The Two Micron All Sky Survey
(2MASS) images of the field from 2000 June 15 {\small UTC} compared
with the three colour images showing the infrared transient (IRT)
several minutes after the GRB triggered. Time relative to the GRB
trigger is given as is the number (N) of individual images
combined. The individual 2MASS images are 1.3 s integrations, so the
combination of 6 2MASS images is equal to a single PAIRITEL image. The
2MASS and PAIRITEL fields are centred on the IRT and are approximately
1\arcmin\ on a side (shown as a blue box in the colour image). A local
background from a 2-D median was subtracted from all images to remove
large-scale background variations. For all images North is up and East
is to the left.  {\bf Right:} The Keck NIRC-1\cite{ms94} imaging at
$J$-band on December 21 {\small UTC}, in the same region as the
PAIRITEL images. NIRC-1 images were combined in the usual manner and
the combined images had a seeing of less than 0.8\arcsec\ on both
nights. The IRT, as well as two unresolved nearby sources (S1, S2),
are labelled. 

\smallskip

{\small {\bf High resolution version of this figure:} {\tt http://astro.berkeley.edu/$\sim$jbloom/grb041219a/fig1.ps}}}
\label{fig:discovery}
\end{figure*}

Prompt long-wavelength afterglow emission is predicted to arise when
the reverse (external) shock encounters the ejecta of the
explosion\cite{mr97a,sp99a}, or through $\gamma$-ray heating of the
circumburst material\cite{bel02}. Indeed, four
GRBs\cite{abb+99,fyk+03,fps+03,lfcj03,rsp+04} have exhibited transient
optical emission that could be associated with reverse shocks, but
early-time optical transients have not been found for the vast
majority of bursts (however, new larger-aperture robotic optical
systems have met with increasing success). Since even moderate levels
of dust near the GRB or along the line-of-sight in the host galaxy
could effectively suppress detectable optical emission\cite{khg+03},
contemporaneous observations at IR wavelengths, where light
suppression is relatively minimised, offer a natural means to uncover
any prompt emission.  This was one motivation for our construction of
the Peters Automated Infrared Imaging Telescope (PAIRITEL; figure
~\ref{fig:discovery}). At 1.3 metres in diameter, it is one of the
largest, completely autonomous telescope systems in the world and one
of only a few capable of imaging at IR wavelengths (1.1--2.3 $\mu$m;
see ref.~\cite{zcg+03}). PAIRITEL acquires images with high temporal
cadence (integration times of 7.8s) in 3 colours simultaneously. The
field of view is rather large, 8.5$^\prime$$\times$8.5$^\prime$, for
IR imaging, allowing for follow-up of GRB localisations of even modest
precision.

\begin{figure*}[tp]
\centerline{\psfig{file=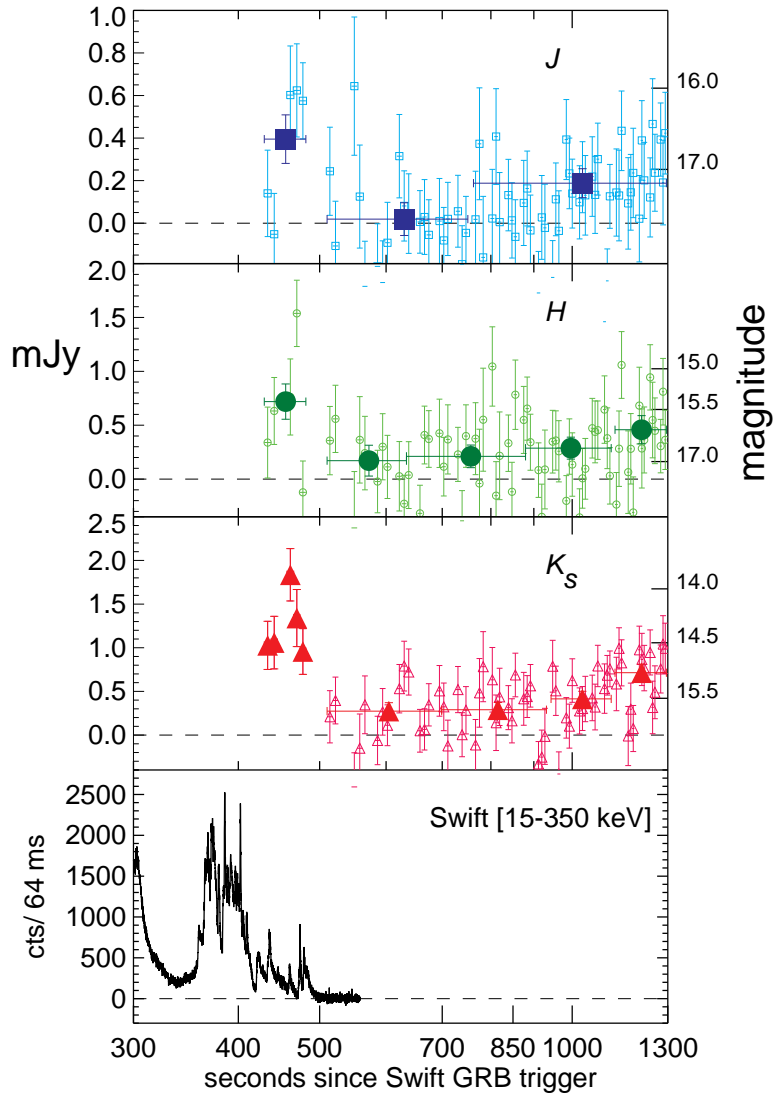,width=4.3in,angle=0}}
\vskip -0.5cm
\caption[] {\small Evolution of the IR flash
associated with GRB\,041219a. We argue that the initial pulse is not
traditional afterglow from an external shock but instead related to
the internal shocks of the central engine. Shown are measurements
derived from individual 7.8 s exposures as small, light-coloured
points, and detections from stacks of images as heavy, dark-coloured
points; error bars are 1 $\sigma$, estimated from photon noise and the
distribution of randomly-placed apertures on the images. Also plotted
is the light curve from the 15--350 keV channels of the BAT instrument
aboard Swift. {\bf Reductions:} The response to variable sky and bias
in the detectors was estimated for each exposure in each band by
median-combining all the exposures taken within a two to four minute
window. A flat-field correction for fixed pixel-to-pixel variations in
detector gain was made using images of the dawn sky. Subtracting the
dark+sky response and normalising by the flat-field produced a reduced
image. {\bf Photometry:} Reduced images were measured either
individually or as stacks of co-added images. Since individual images
undersample the seeing due to atmospheric blurring, photometry was
performed differentially in an aperture of fixed size.  All of the
images were aligned to a common reference image to an accuracy of
approximately 0.1 pixel. Images were measured individually or in
stacks created by summing individual images with weights determined by
the signal-to-noise of each image. The measured flux at the position
of the GRB was compared to the flux measured for a set of nearby
comparison stars. Magnitudes of the comparison stars are known to a
high accuracy ($2$\%) from the 2MASS catalog.  PAIRITEL uses the same
detectors and filters as 2MASS, so our differential photometry is
expected to be free from systematic offsets.}
\label{fig:photom}
\end{figure*}

GRB\,041219a triggered the IBIS instrument on-board the {\it
INTEGRAL}\cite{Winkler} satellite on 2004 December 19 at 01:42:54
{\small UTC} which was reported at 01:44:05 {\small UTC}. The initial
position of right ascension (RA) 00h 24min 26s, declination (dec.)
+62\degree\ 50\arcmin\ 06\arcsec\ was refined to 2\arcmin\ uncertainty
at 01:47:49 {\small UTC} and a final offline location\cite{gms+04} was
reported at 03:31:58 {\small UTC}. The Burst Alert Telescope (BAT) on
the Swift satellite\cite{gcm+04} triggered and located GRB\,041219a
on-board at 01:42:18 {\small UTC} with a position that was within
4\arcmin\ of the IR source. If Swift had not still been in its
commissioning phase with slewing disabled, the spacecraft could have
slewed to the location within 70s of the BAT trigger.  The groundbased
BAT location (RA 00h 24m 37.0s, dec.\ +62d 50\arcmin 49.2\arcsec) was
within 48\arcsec\ of the IR source. As viewed by BAT, the burst
duration ($\Delta t$) above background was $520$s and was very bright,
with up to $6.5 \times 10^{4}$ cts/sec (unsaturated) between 15 and
350 keV and a fluence of $1.15 \times 10^{-4}$ erg cm$^{-2}$. The time
evolution of the count rate in four BAT channels covering 15 to 350
keV is reproduced in the lowest panel of figure \ref{fig:photom}.

PAIRITEL began to slew on 2004 December 19 01:48:20 {\small UTC} and
the first observations of the GRB field commenced 58 seconds
later. Despite very poor observing conditions (sustained 40 mph winds,
variable sky transmission, and 4\arcsec\ seeing), comparison of the
first epoch of data revealed a new, variable source\cite{bb04} not
visible in the 2MASS catalogue images of the field.  When compared to
the astrometric grid of 2MASS stars in the field, we find the absolute
position of the IR transient (IRT) to be RA 00h 24min 27.68s $\pm$
0.124\arcsec, dec.~+62$^\circ$ 50\arcmin\ 33.501\arcsec\ $\pm$
0.228\arcsec, with its uncertainty dominated by the mapping to 2MASS
catalogue stars.  An optical flash was also detected in 10 images
acquired before $t$+400 s by the RAPTOR experiment during the prompt
$\gamma$-ray emission\cite{wve+04}.

PAIRITEL observations of the transient continued over the following
three nights, until inclement weather in Arizona precluded additional
observations. In total, 5,790 images were acquired by the system over
these nights. In addition, we obtained deep $J$-band imaging on
December 20 and 21 {\small UTC} using the NIRC-1\cite{ms94} instrument
on the Keck I 10 metre telescope on Mauna Kea, Hawaii. Consistent
within the astrometric accuracy of the IRT from December 19, we found
a point-like source in $J$-band (figure \ref{fig:discovery}). Over
these two nights, that source was seen to fade by 1.0 magnitudes,
confirming its identification with the IRT. The Keck images revealed
two sources within 2.5\arcsec\ of the transient position (S1:
$J\approx 19.7$\,mag, 2.5\arcsec\ North-north-east; S2: $J \approx
21.4$\,mag, 1.5\arcsec\ East): both were unresolved apparent point
sources. The source S1 is bright enough to contaminate the PAIRITEL
$J$-band aperture photometry on Dec 21, which accounts for the
difference between our measurements and the fainter measurements from
Apache Point\cite{apo} and Keck on the same date. When comparing
PAIRITEL photometry with higher-resolution Keck and APO results for
Dec 21, the flux from S1+S2 appears to be a {51\%} contamination in
$H$-band, {58\%} in $J$ band, and has negligible contribution in the
${K_s}$ band.

The resulting light curves (see Supplemental Table \#1) shown in
figure\ \ref{fig:long} reveal a complex time history of the
afterglow. The first six PAIRITEL exposures at $t$+7.2 min after the
trigger show a source that brightens, then fades very rapidly in all
filters by about $t$+9 min, and then rebrightens by $t$+20 min. Using
our data and data reported in the literature, we fitted the light
curves as the sum of three smoothly-connected rise and fall
brightening events; the results of these fits are shown in
figure~\ref{fig:long}. During the first few days, the source
colours, though rather uncertain, appear consistent with a single
value of the spectral slope of $\beta\approx0.4$ (figure
4). Additionally, there is some evidence that the IRT was redder
during the ``flash'' event at $t$+7.2 min.

How might the light curve be understood as emission from the reverse
and forward shock? The electrons in the shock are assumed to be
accelerated to a power-law spectrum with number density as a function
of energy ($E$) proportional to $E^{-p}$.  In a constant-density
circumburst environment, a reverse shock is expected (see
ref.~\cite{kob00}) to rise rapidly ($\alpha = 3p - 3/2$, with $f_\nu
\propto t^\alpha$) and then, in the ``thin shell'' case (see below),
decline with $\alpha = -(27p + 7)/35 \approx -2$ after the emission
peak, corresponding to the time ($t_{\times}$) that the reverse shock
crosses the explosion ejecta. The IR emission from the forward
shock\cite{spn98} should rise slowly as $\alpha = 0.5$, then decline
as $\alpha = 3(1 - p)/4$.  Associating peak 2 with the reverse shock
and peak 3 with the forward shock, we find reasonable agreement, to
within the measurement uncertainties, of the data with this model. In
particular, the three implied values for $p$ ($2.5 \pm 1.1$, $4.2 \pm
4.1$, $2.6 \pm 0.2$, for the reverse-rise, reverse-decline, and
forward-decline, respectively) are all consistent with the usual
range\cite{vkw00} of $p=2.2-2.5$. A consequence of this interpretation
is that, in the absence of effects due to collimation of the burst,
radio afterglow emission should be dominated by a rising reverse
shock, peaking at time $t = (10.2 \pm 1.8) (\nu/8.4~{\rm
GHz})^{-35/54}$ day; this is thus far confirmed with reports of rising
radio emission at least to day 2.9 (see \cite{vdh+04}).

With this interpretation, there are three puzzles. First, we would
expect\cite{sp99a} the source to have become bluer during the forward
shock rise, which is not required (though is not excluded) by our
data. The second puzzle concerns the timing of the reverse shock peak
relative to the GRB duration. We expect\cite{sar97,zkm03} $t_{\times}
= 1670$s $ > \Delta t$ only when $\Delta t$ is less than the time when
the shock begins to decelerate (commonly deemed the ``thin shell''
case). This deceleration time occurs when the shock has swept up from
the circumburst environment a quantity $1/\Gamma_0$ times the
entrained mass, where $\Gamma_0$ is the terminal Lorentz factor of the
shock. A delayed reverse shock crossing time requires $\Gamma_0 < 67
E_{52}^{1/8} n^{-1/8} (\Delta t/520{\rm s})^{-3/8} [0.5 (1+z)]^{3/8}$,
where $n$ is the circumburst particle density in units of baryons
cm$^{-3}$, $E_{52}$ is the energy in the shock in units of $10^{52}$
erg, and the redshift of the burst is $z$. This initial Lorentz factor
constraint is uncomfortably smaller than the limits placed on previous
values\cite{zkm03} of $\Gamma_0$, which suggests that either the burst
occurred at high redshift (which is excluded by the RAPTOR
detection\cite{wve+04}), was exceptionally energetic, or occurred in a
low-density environment. Alternatively, the observed variability could
be due to inhomogeneous density structure in the circumburst
environment or delayed energy injection into the blastwave.

The third and most intriguing puzzle is the physical origin of the IR
flash at $t1$ = 462s that has marginal evidence for appearing more red
than the afterglow at later times. Regardless of the interpretation of
events 2 and 3, the rapid rise and fall appear to preclude an
association with a reverse shock: the fit of a power-law decay slope
to the $K_s$ data acquired less than 10 min from burst trigger yields
$\alpha = -18 \pm 5$ whereas setting $\alpha=-2$, as expected of a
reverse shock, yields an unacceptable fit (reduced $\chi^2 = 4.4$).
The duration of the IR flash ($\delta t \approx 45$s, taken as the
full-width at half-maximum of the model fit to the data) is comparable
to the widths of the largest timescale for substructure in
$\gamma$-rays; indeed, this $\delta t$ is remarkably similar to that
of the optical pulse\cite{abb+99} in GRB\,990123. The ratio of the
width to the time after trigger $t \sim 462$s, $\delta t/t = 0.10$, is
similar to that seen in individual pulses in bright GRBs\cite{rf00}
but a factor of $\sim$10 smaller than the $\delta t/t$ of the optical
flash of GRB\,990123.

We suggest, therefore, that the origin of the first peak is from the
internal shock that itself produced the GRB. Such emission is possible
if the synchrotron cooling frequency from the internal shock emission
is well below $\gamma$-ray frequencies\cite{sp99a}.  Indeed, if the IR
emission is due to internal shocks, then the observed flash may be due
to a superposition of several unresolved shorter-timescale
pulses. Future observations of IR flashes in the Swift era will no
doubt test the ubiquity and nature of rapidly variable early-time
emission as presented herein.

\vskip 0.7in

\begin{thebibliography}{10}
\expandafter\ifx\csname url\endcsname\relax
  \def\url#1{\texttt{#1}}\fi
\expandafter\ifx\csname urlprefix\endcsname\relax\def\urlprefix{URL }\fi
\providecommand{\bibinfo}[2]{#2}
\providecommand{\eprint}[2][]{\url{#2}}

\bibitem{fmn96}
\bibinfo{author}{{Fenimore}, E.~E.}, \bibinfo{author}{{Madras}, C.~D.} \&
  \bibinfo{author}{{Nayakshin}, S.}
\newblock \bibinfo{title}{{Expanding Relativistic Shells and Gamma-Ray Burst
  Temporal Structure}}.
\newblock \emph{\bibinfo{journal}{Astrophys. J.}}
  \textbf{\bibinfo{volume}{473}}, \bibinfo{pages}{998--+}
  (\bibinfo{year}{1996}).

\bibitem{mr97a}
\bibinfo{author}{M\'esz\'aros, P.} \& \bibinfo{author}{Rees, M.~J.}
\newblock \bibinfo{title}{Optical and long-wavelength afterglow from gamma-ray
  bursts}.
\newblock \emph{\bibinfo{journal}{Astrophys. J.}}
  \textbf{\bibinfo{volume}{476}}, \bibinfo{pages}{232--237}
  (\bibinfo{year}{1997}).

\bibitem{sp99a}
\bibinfo{author}{{Sari}, R.} \& \bibinfo{author}{{Piran}, T.}
\newblock \bibinfo{title}{Predictions for the very early afterglow and the
  optical flash}.
\newblock \emph{\bibinfo{journal}{Astrophys. J.}}
  \textbf{\bibinfo{volume}{520}}, \bibinfo{pages}{641--649}
  (\bibinfo{year}{1999}).

\bibitem{bel02}
\bibinfo{author}{{Beloborodov}, A.~M.}
\newblock \bibinfo{title}{{Radiation Front Sweeping the Ambient Medium of
  Gamma-Ray Bursts}}.
\newblock \emph{\bibinfo{journal}{Astrophys. J.}}
  \textbf{\bibinfo{volume}{565}}, \bibinfo{pages}{808--828}
  (\bibinfo{year}{2002}).

\bibitem{vkw00}
\bibinfo{author}{{van Paradijs}, J.}, \bibinfo{author}{{Kouveliotou}, C.} \&
  \bibinfo{author}{{Wijers}, R.~A.~M.~J.}
\newblock \bibinfo{title}{{Gamma-Ray Burst Afterglows}}.
\newblock \emph{\bibinfo{journal}{Ann. Rev. Astr. Ap.}}
  \textbf{\bibinfo{volume}{38}}, \bibinfo{pages}{379--425}
  (\bibinfo{year}{2000}).

\bibitem{bb04a}
\bibinfo{author}{{Baring}, M.~G.} \& \bibinfo{author}{{Braby}, M.~L.}
\newblock \bibinfo{title}{{A Study of Prompt Emission Mechanisms in Gamma-Ray
  Bursts}}.
\newblock \emph{\bibinfo{journal}{Astrophys. J.}}
  \textbf{\bibinfo{volume}{613}}, \bibinfo{pages}{460--476}
  (\bibinfo{year}{2004}).

\bibitem{abb+99}
\bibinfo{author}{{Akerlof}, C.} \emph{et~al.}
\newblock \bibinfo{title}{Observation of contemporaneous optical radiation from
  a gamma-ray burst.}
\newblock \emph{\bibinfo{journal}{Nature}} \textbf{\bibinfo{volume}{398}},
  \bibinfo{pages}{400--402} (\bibinfo{year}{1999}).

\bibitem{gms+04}
\bibinfo{author}{{Gotz}, D.}, \bibinfo{author}{{Mereghetti}, S.},
  \bibinfo{author}{{Shaw}, S.}, \bibinfo{author}{{Beck}, M.} \&
  \bibinfo{author}{{Borkowski}, J.}
\newblock \bibinfo{title}{{GRB 041219 - A long GRB detected by INTEGRAL}}.
\newblock \emph{\bibinfo{journal}{{GCN Circular}}}
  \textbf{\bibinfo{volume}{2866}}, \bibinfo{pages}{}
  (\bibinfo{year}{2004}).

\bibitem{fyk+03}
\bibinfo{author}{{Fox}, D.~W.} \emph{et~al.}
\newblock \bibinfo{title}{{Early optical emission from the {$\gamma$}-ray burst
  of 4 October 2002}}.
\newblock \emph{\bibinfo{journal}{Nature}} \textbf{\bibinfo{volume}{422}},
  \bibinfo{pages}{284--286} (\bibinfo{year}{2003}).

\bibitem{fps+03}
\bibinfo{author}{{Fox}, D.~W.} \emph{et~al.}
\newblock \bibinfo{title}{{Discovery of Early Optical Emission from GRB
  021211}}.
\newblock \emph{\bibinfo{journal}{Astrophys. J.}}
  \textbf{\bibinfo{volume}{586}}, \bibinfo{pages}{L5--L8}
  (\bibinfo{year}{2003}).

\bibitem{lfcj03}
\bibinfo{author}{{Li}, W.}, \bibinfo{author}{{Filippenko}, A.~V.},
  \bibinfo{author}{{Chornock}, R.} \& \bibinfo{author}{{Jha}, S.}
\newblock \bibinfo{title}{{The Early Light Curve of the Optical Afterglow of
  GRB 021211}}.
\newblock \emph{\bibinfo{journal}{Astrophys. J.}}
  \textbf{\bibinfo{volume}{586}}, \bibinfo{pages}{L9--L12}
  (\bibinfo{year}{2003}).

\bibitem{rsp+04}
\bibinfo{author}{{Rykoff}, E.~S.} \emph{et~al.}
\newblock \bibinfo{title}{{The Early Optical Afterglow of GRB 030418 and
  Progenitor Mass Loss}}.
\newblock \emph{\bibinfo{journal}{Astrophys. J.}}
  \textbf{\bibinfo{volume}{601}}, \bibinfo{pages}{1013--1018}
  (\bibinfo{year}{2004}).

\bibitem{khg+03}
\bibinfo{author}{{Klose}, S.} \emph{et~al.}
\newblock \bibinfo{title}{{The Very Faint K-Band Afterglow of GRB 020819 and
  the Dust Extinction Hypothesis of the Dark Bursts}}.
\newblock \emph{\bibinfo{journal}{Astrophys. J.}}
  \textbf{\bibinfo{volume}{592}}, \bibinfo{pages}{1025--1034}
  (\bibinfo{year}{2003}).

\bibitem{zcg+03}
\bibinfo{author}{{Zerbi}, F.~M.} \emph{et~al.}
\newblock \bibinfo{title}{{REM telescope, a robotic facility to monitor the
  prompt afterglow of Gamma Ray Bursts}}.
\newblock In \bibinfo{editor}{Iye, M.} \& \bibinfo{editor}{Moorwood, A. F.~M.}
  (eds.) \emph{\bibinfo{booktitle}{Instrument Design and Performance for
  Optical/Infrared Ground-based Telescopes. Proceedings of the SPIE.}}, vol.
  \bibinfo{volume}{4841}, \bibinfo{pages}{737--748} (\bibinfo{year}{2003}).

\bibitem{Winkler}
\bibinfo{author}{{Winkler}, C.~e.}
\newblock \bibinfo{title}{{The INTEGRAL Mission}}.
\newblock \emph{\bibinfo{journal}{AA}} \textbf{\bibinfo{volume}{411}},
  \bibinfo{pages}{1--6} (\bibinfo{year}{2003}).

\bibitem{gcm+04}
\bibinfo{author}{{Gehrels}, N.} \emph{et~al.}
\newblock \bibinfo{title}{{The Swift Gamma-Ray Burst Mission}}.
\newblock \emph{\bibinfo{journal}{Astrophys. J.}}
  \textbf{\bibinfo{volume}{611}}, \bibinfo{pages}{1005--1020}
  (\bibinfo{year}{2004}).

\bibitem{bb04}
\bibinfo{author}{{Blake}, C.} \& \bibinfo{author}{{Bloom}, J.~S.}
\newblock \bibinfo{title}{{GRB 041219: Infrared Afterglow Candidate}}.
\newblock \emph{\bibinfo{journal}{{GCN Circular}}}
  \textbf{\bibinfo{volume}{2870}}, \bibinfo{pages}{}
  (\bibinfo{year}{2004}).

\bibitem{wve+04}
\bibinfo{author}{{Vestrand}} \emph{et~al.}
\newblock \bibinfo{title}{{A Link between Prompt Optical and Prompt Gamma-Ray
  Emission in Gamma-Ray Bursts}}.
\newblock \emph{\bibinfo{journal}{Nature}} \textbf{\bibinfo{volume}{{?}}},
  \bibinfo{pages}{?--+} (\bibinfo{year}{2005}).

\bibitem{ms94}
\bibinfo{author}{Matthews, K.} \& \bibinfo{author}{Soifer, B.~T.}
\newblock \bibinfo{title}{{NIRC} infrared instrument}.
\newblock In \bibinfo{editor}{McLean, I.} (ed.)
  \emph{\bibinfo{booktitle}{Infrared Astronomy with Arrays, the Next
  Generation}}, \bibinfo{pages}{239} (\bibinfo{publisher}{Dordrecht: Kluwer},
  \bibinfo{year}{1994}).

\bibitem{apo}
\bibinfo{author}{{Heary}, F.}
\newblock \bibinfo{title}{{NIR Observations of GRB 041219}}.
\newblock \emph{\bibinfo{journal}{{GCN Circular}}}
  \textbf{\bibinfo{volume}{2916}}, \bibinfo{pages}{}
  (\bibinfo{year}{2004}).

\bibitem{kob00}
\bibinfo{author}{{Kobayashi}, S.}
\newblock \bibinfo{title}{{Light Curves of Gamma-Ray Burst Optical Flashes}}.
\newblock \emph{\bibinfo{journal}{Astrophys. J.}}
  \textbf{\bibinfo{volume}{545}}, \bibinfo{pages}{807--812}
  (\bibinfo{year}{2000}).

\bibitem{spn98}
\bibinfo{author}{{Sari}, R.}, \bibinfo{author}{{Piran}, T.} \&
  \bibinfo{author}{{Narayan}, R.}
\newblock \bibinfo{title}{{Spectra and Light Curves of Gamma-Ray Burst
  Afterglows}}.
\newblock \emph{\bibinfo{journal}{Astrophys. J.}}
  \textbf{\bibinfo{volume}{497}}, \bibinfo{pages}{L17} (\bibinfo{year}{1998}).

\bibitem{vdh+04}
\bibinfo{author}{van~der Horst, A.}, \bibinfo{author}{Rol, E.} \&
  \bibinfo{author}{Strom, R.}
\newblock \bibinfo{title}{{GRB 041219: Second Epoch WSRT Radio Observations}}.
\newblock \emph{\bibinfo{journal}{{GCN Circular}}}
  \textbf{\bibinfo{volume}{2895}}, \bibinfo{pages}{}
  (\bibinfo{year}{2004}).

\bibitem{sar97}
\bibinfo{author}{Sari, R.}
\newblock \bibinfo{title}{Hydrodynamics of gamma-ray burst afterglow}.
\newblock \emph{\bibinfo{journal}{Astrophys. J.}}
  \textbf{\bibinfo{volume}{489}}, \bibinfo{pages}{L37--L40}
  (\bibinfo{year}{1997}).

\bibitem{zkm03}
\bibinfo{author}{{Zhang}, B.}, \bibinfo{author}{{Kobayashi}, S.} \&
  \bibinfo{author}{{M{\' e}sz{\' a}ros}, P.}
\newblock \bibinfo{title}{{Gamma-Ray Burst Early Optical Afterglows:
  Implications for the Initial Lorentz Factor and the Central Engine}}.
\newblock \emph{\bibinfo{journal}{Astrophys. J.}}
  \textbf{\bibinfo{volume}{595}}, \bibinfo{pages}{950--954}
  (\bibinfo{year}{2003}).

\bibitem{rf00}
\bibinfo{author}{{Ramirez-Ruiz}, E.} \& \bibinfo{author}{{Fenimore}, E.~E.}
\newblock \bibinfo{title}{{Pulse Width Evolution in Gamma-Ray Bursts: Evidence
  for Internal Shocks}}.
\newblock \emph{\bibinfo{journal}{Astrophys. J.}}
  \textbf{\bibinfo{volume}{539}}, \bibinfo{pages}{712--717}
  (\bibinfo{year}{2000}).

\bibitem{np04}
\bibinfo{author}{{Nakar}, E.} \& \bibinfo{author}{{Piran}, T.}
\newblock \bibinfo{title}{{Early afterglow emission from a reverse shock as a
  diagnostic tool for gamma-ray burst outflows}}.
\newblock \emph{\bibinfo{journal}{Mon. Not. R. astr. Soc.}}
  \textbf{\bibinfo{volume}{353}}, \bibinfo{pages}{647--653}
  (\bibinfo{year}{2004}).

\bibitem{moon2}
\bibinfo{author}{{Moon}, D.-S.}, \bibinfo{author}{{Cenko}, B.} \&
  \bibinfo{author}{{Adams}, J.}
\newblock \bibinfo{title}{{GRB041219: Continued NIR Observations}}.
\newblock \emph{\bibinfo{journal}{{GCN Circular}}}
  \textbf{\bibinfo{volume}{2884}}, \bibinfo{pages}{}
  (\bibinfo{year}{2004}).

\bibitem{sfd98}
\bibinfo{author}{Schlegel, D.~J.}, \bibinfo{author}{Finkbeiner, D.~P.} \&
  \bibinfo{author}{Davis, M.}
\newblock \bibinfo{title}{Maps of dust infrared emission for use in estimation
  of reddening and cosmic microwave background radiation foregrounds}.
\newblock \emph{\bibinfo{journal}{Astrophys. J.}}
  \textbf{\bibinfo{volume}{500}}, \bibinfo{pages}{525--553}
  (\bibinfo{year}{1998}).

\bibitem{zband}
\bibinfo{author}{Cenko, S.~B.}
\newblock \bibinfo{title}{{GRB 041219: Optical Afterglow Detection}}.
\newblock \emph{\bibinfo{journal}{{GCN Circular}}}
  \textbf{\bibinfo{volume}{2885}}, \bibinfo{pages}{}
  (\bibinfo{year}{2004}).

\end{thebibliography}

\begin{addendum}

 \item[Correspondence] Correspondence and requests for materials
should be addressed to J.S.B.\ \\(email: jbloom@astro.berkeley.edu).

 \item J.S.B. was supported by a Junior Fellowship from the Harvard
 Society of Fellows.  PAIRITEL was made possible by a grant from the
 Harvard Milton Fund. Additional funding from the Smithsonian
 Institution for the PAIRITEL project is gratefully acknowledged. We
 would like to thank the entire Mt.~Hopkins Ridge Staff for support of
 PAIRITEL, especially W.\ Peters, R.  Hutchins, and T. ~Groner.  J.\
 Huchra is thanked for stewardship over the telescope under adverse
 conditions in the week leading up to GRB\,041219a. This publication
 makes use of data products from the Two Micron All Sky Survey
 (2MASS), which is a joint project of the University of Massachusetts
 and the Infrared Processing and Analysis Center/California Institute
 of Technology, funded by the National Aeronautics and Space
 Administration and the National Science Foundation.

 \item[Competing Interests] The authors declare that they have no
competing financial interests.

\end{addendum}

\begin{figure*}[tp]
\centerline{\psfig{file=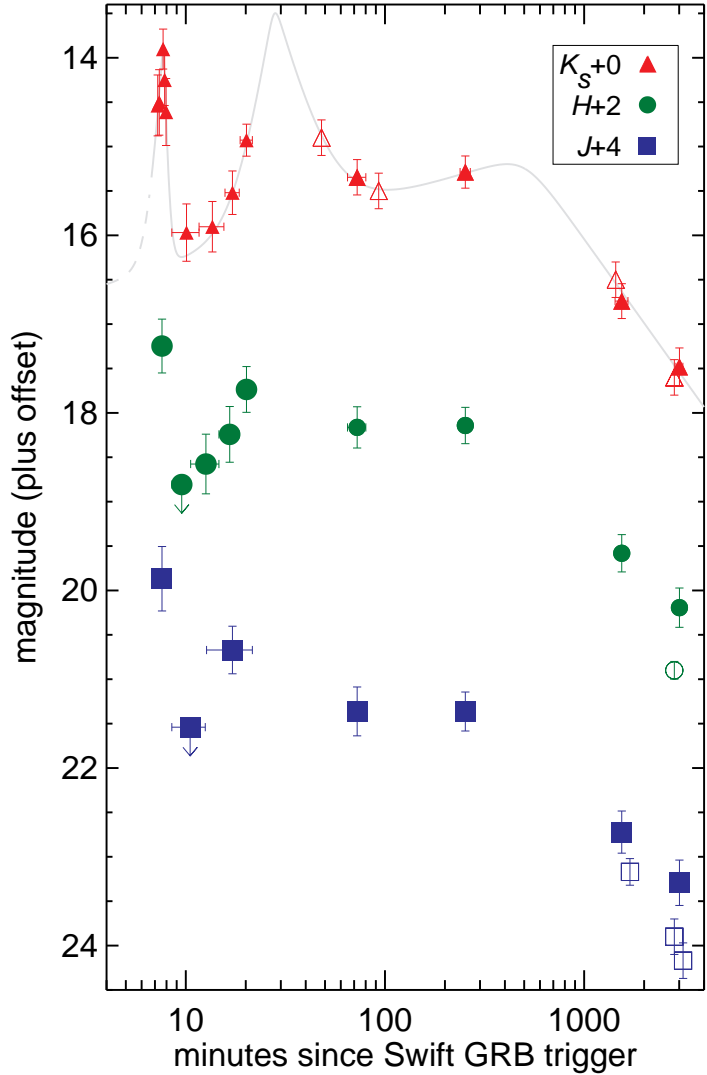,width=4.3in,angle=0}}
\caption[] {\small Long-timescale variability of the IR transient
associated with GRB\,041219a. The $K_s$-band light-curve is reasonably
fit (grey curve) as a sum of three power-law rise and fall events with
peaks at times since the burst $[t1,t2,t3]$ = [$7.73$m $\pm$ 0.14m,
27.5m $\pm 4.9$m, $\approx$500m] and specific brightness
[$f_\nu(t1),f_\nu(t2),f_\nu(t3)$] = [1.6 $\pm$ 0.3 mJy, 2.5 $\pm$ 2.8
mJy, 0.5 $\pm$ 0.1 mJy]. The rise and fall near the second peak is
fitted by $\alpha_{2,r} = 6.1 \pm 2.9$ ($\alpha_{2,f} = -3.4 \pm
2.8$), with the specific brightness changing as $f_\nu
\propto t^\alpha$. The final rise and fall parameters depend on the
poorly constrained value of $t_3$. Fixing $t_3 = 500$m, we find
$\alpha_{3,r} = 0.3$ $\pm$ 0.1 and $\alpha_{3,f} = -1.2 \pm 0.1$. The
first peak has both rapid rise and fall times (see text). The global
$\chi^2$ per degree of freedom (dof) is an acceptable 1.24. The
reported 1-$\sigma$ parameter uncertainties do not reflect the
covariance between the parameter fits. Additionally, all of these
derived values depend on the choice of a smoothing parameter to
connect the two power-laws ($s$ in eq.~2 of Nakar \& Piran\cite{np04};
here, we chose $s=5$). The $J$- and $H$-band light curves are
consistent with the $K_s$-band shape, aside from the overall flux
normalisation.  {\bf Photometry}: Large numbers of individual
exposures were combined with sub-pixel sampling and produced a high
signal-to-noise, high-resolution image for each epoch.  The
point-spread function in these images is well-sampled, so the
technique of difference image photometry was utilised.  The
uncertainty in the differential fluxes of the IRT between individual
epochs is near photon-limited.  The differential fluxes are converted
to magnitudes from aperture photometry at the position of the IRT in
the late-time (Dec 21) reference image.  Other data, plotted as open
points, include two $J$-band observations made with the Keck I
telescope, $JHK_{s}$ observations from Apache Point
Observatory\cite{apo}, and three $K_{s}$ measurements (see
\cite{moon2} and references therein) from Mt.\ Palomar.}
\label{fig:long}
\end{figure*}

\begin{figure*}[tp]		
\centerline{\psfig{file=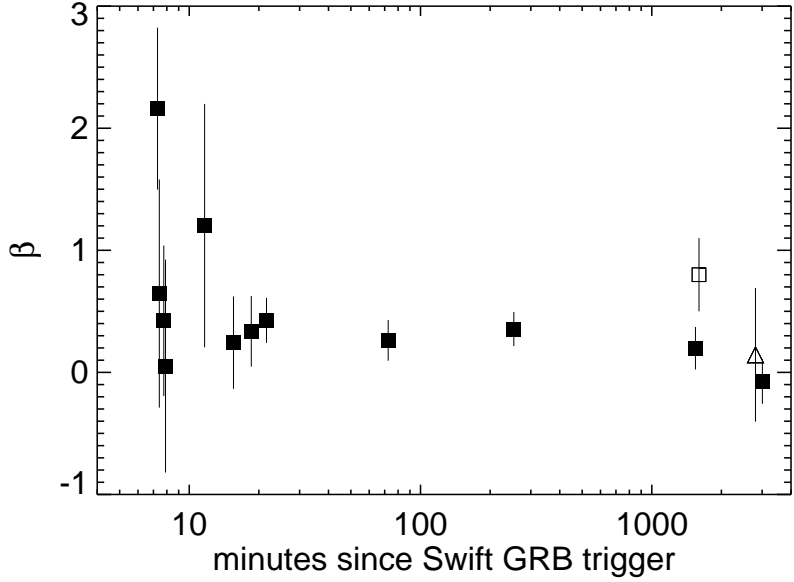,width=4.8in,angle=0}}
\caption[]{\small Colour evolution of the afterglow of GRB\,041219a,
corrected for extinction due to interstellar dust. The dust
extinction\cite{sfd98} at the location of GRB 041219a is high
\hbox{$A(V$ band$)$} = 5.9 (and highly uncertain). Using a common dust
extinction law ($R_V = 3.1$), the extinction in magnitudes is $A(J)$ =
1.6, $A(H)$ = 1.04, and $A(K_s)$ = 0.66. We fitted (solid squares) for
a spectral index $\beta$ with our $J$, $H$, and $K_{s}$ measurements
at each epoch using the flux relation
$F_\nu(t)=F_{0}(t)(\lambda_{c}/\lambda_{H})^{\beta}$, where
$\lambda_{c}$ is the central wavelength of the filter bandpasses ($c =
[J, H, K]$).  Including the $z$-band detection\cite{zband} ($\lambda_c
\approx 9100$ \AA) that was nearly coeval with our IR observations,
results in a steeper spectral slope of $0.80 \pm 0.39$ (open square)
than found with the IR data alone. Our fit to the Apache Point $JHK$
data\cite{apo} is also shown with an open triangle. With fits that
yielded $\chi^2$/dof greater than unity, we scale the resulting 1
$\sigma$ errors by $\sqrt{\chi^2}$. Though there is some evidence for
secular trends, the resulting values for $\beta$ are consistent with a
single value of $\beta \approx 0.4$. We caution, however, that this
common value of the spectral index is strongly dependent upon the
assumed dust column. For different values of the extinction, $A(V)$ =
4.0 and $A(V)$ = 7.0, the fits to the spectral index are $\beta
\approx 0.7$ and $\beta \approx 0.2$, respectively. Indeed, the
standard forward-shock model\cite{spn98} tends to favour a larger
value ($\beta \approx 0.55$) than found with the IR data alone.  For
the first peak (during the GRB), there is marginal evidence that the
source is intrinsically redder than at later times.}
\label{fig:colour}
\end{figure*}

\end{document}